\documentclass[a4paper,11pt]{article}

\usepackage{jheppub} 

\usepackage[T1]{fontenc} 
\usepackage{tabularx}

\title{CP violating effects in $^{210}$Fr and prospects for new physics beyond the Standard Model
}

\author[a,b]{Nanako Shitara,}
\author[c,d,e]{Nodoka Yamanaka,}
\author[f]{Bijaya Kumar Sahoo,}
\author[g]{Toshio Watanabe}
\author[a,h]{and Bhanu Pratap Das}

\affiliation[a]{Department of Physics, School of Science, Tokyo Institute of Technology, Ookayama, Meguro-ku, Tokyo 152-8550, Japan}
\affiliation[b]{Department of Physics, University of Colorado, Boulder, CO 80309, U.S.A.}
\affiliation[c]{Amherst Center for Fundamental Interactions, Department of Physics, University of Massachusetts Amherst, MA 01003, U.S.A.}
\affiliation[d]{Department of Physics, Kennesaw State University, Kennesaw, GA 30144, U.S.A.}
\affiliation[e]{Nishina Center for Accelerator-Based Science, RIKEN, Wako 351-0198, Japan}
\affiliation[f]{Atomic, Molecular and Optical Physics Division, Physical Research Laboratory, Navrangpura, Ahmedabad 380009, India}
\affiliation[g]{Global Scientific Information and Computing Center, Tokyo Institute of Technology, Ookayama, Meguro-ku, Tokyo 152-8550, Japan}
\affiliation[h]{Centre for Quantum Engineering Research and Engineering, TCG CREST, Sector V, Salt Lake, Kolkata 700091, India}

\emailAdd{nanako.shitara@colorado.edu}
\emailAdd{nodoka.yamanaka@riken.jp}
\emailAdd{bijaya@prl.res.in}
\emailAdd{t.watanabe@gsic.titech.ac.jp}
\emailAdd{das.b.aa@m.titech.ac.jp}

\abstract{
We report theoretical results of the electric dipole moment (EDM) of $^{210}$Fr which arises from the interaction of the EDM of an electron with 
the internal electric field in an atom and the scalar-pseudoscalar electron-nucleus interaction; the two dominant sources of CP violation in this
atom. Employing the relativistic coupled-cluster theory, we evaluate the enhancement factors for these two CP violating interactions to an accuracy 
of about 3\% and analyze the contributions of the many-body effects. These two quantities in combination with the projected sensitivity of the 
$^{210}$Fr EDM experiment provide constraints on new physics beyond the Standard Model. Particularly, we demonstrate that their precise values are necessary to account for the effect of the bottom quark in 
models in which the Higgs sector is augmented by nonstandard Yukawa interactions such as the two-Higgs doublet model.
}

\begin{document} 
\maketitle
\flushbottom

\section{Introduction}

For all its successes in explaining a wide range of phenomena involving fundamental particles, the Standard Model (SM) of particle physics is unable 
to account for some outstanding observations in the universe. One of these is the so-called baryon number asymmetry, where the matter-antimatter ratio
observed in the current universe differs by several orders of magnitude from the value predicted by the SM~\cite{Canetti}. One of the possible reasons 
for this discrepancy could be the existence of additional sources of CP-violation, which is the combined violation of charge conjugation (C) and parity
(P) symmetries~\cite{Canetti}. A signature of CP-violation that has not yet been observed is the intrinsic electric dipole moment (EDM) of the electron. 
For a particle to possess such an EDM, both P and time-reversal (T) symmetries must be independently violated~\cite{Landau, Ballentine, Sandars1,
Kriplovich}. From the CPT theorem, T-violation implies CP-violation. The SM predicts an electron EDM (eEDM) value, denoted by $d_e$, of $\vert d_e 
\vert \sim 10^{-39}e$\, cm~\cite{Yamaguchi:2020eub,Yamaguchi:2020dsy}, which falls far outside the range of values current experiments can probe. 
However, several beyond the SM (BSM) paradigms predict values of the eEDM that are many orders of magnitude larger, such as variants of the 
Supersymmetric (SUSY) model and the left-right symmetric model, which, depending on the parameters, can predict a value of the eEDM as large as 
$\vert d_e \vert \sim10^{-27} e$\, cm~\cite{Bernreuther, Cirigliano}. This is within the reach of the current experiments. Thus, a successful 
measurement of a nonzero eEDM would provide direct evidence for BSM physics~\cite{Cesarotti}, as well as shed insights into the observed baryon 
number asymmetry of the universe~\cite{Fuyuto}. Even if the eEDM is not observed, imposing upper limits on its magnitude can constrain BSM models,
which predict different ranges of the possible eEDM values~\cite{Fukuyama,Roussy:2020ily}.

The observation of the eEDM has eluded experimentalists for over half a century, but not without significant improvements to the upper limits. 
Currently, heavy open-shell atoms and polar molecules are the most promising systems for determining the upper bounds on the magnitude of the eEDM, 
with the best limit to date set by experiments on thorium oxide (ThO), at $\vert d_e \vert < 1.1\times10^{-29} e$\, cm with 90\% confidence~\cite{ACME}.
The best experimental limit using atoms comes from $^{205}$Tl, at $\vert d_e \vert \leq 1.6 \times 10^{-27} e$\, cm with 90\% confidence~\cite{Regan}. 
Because the eEDM is known to be extremely small, high precision is required for these experiments. In the case of atomic systems, the presence of a 
permanent eEDM can induce an atomic EDM ($D_a$) which can be many times larger than the magnitude of the eEDM~\cite{Sandars2}. It is this enhanced 
EDM that is exploited in eEDM experiments using atoms. To obtain an upper limit for the eEDM, the enhancement factor $R=D_a/d_e$, defined as the ratio
of the atomic EDM to the eEDM, must be theoretically evaluated. 

Paramagnetic systems like the ones mentioned above are also sensitive to CP-violation due to the scalar-pseudoscalar (S-PS) interaction between the 
electrons and the nucleons (e-N), which arises from CP violating BSM particle exchanges between the electron and quarks or gluons at the fundamental 
level. This leads to a finite value of the e-N S-PS coupling coefficient ($C_N^{S-PS}$) which is generated in certain extensions of the SM and may be 
probed with the measurement of atomic EDMs. Analogous to $R$ for the electron EDM, for the S-PS e-N interaction, we define $S$, and its evaluation
also requires atomic many-body theory. The S-PS e-N interaction Hamiltonian in atomic systems scales as $Z^3$, where $Z$ is the atomic number. Thus 
Fr being the heaviest alkali atom, is well suited for providing insights into the S-PS e-N interaction. 

In this work, we calculate $R$ and $S$ for $^{210}$Fr in the ground state. In many respects, Fr is a suitable candidate for an EDM experiment. An 
electron EDM search experiment using $^{211}$Fr has been proposed by Wundt et al.~\cite{Wundt} and Munger et al.~\cite{Munger}, at TRIUMF in Canada 
(there are no significant differences between the $R$ as well as the $S$ values for $^{210}$Fr and $^{211}$Fr). Another eEDM search experiment 
using Fr is currently in progress at the University of Tokyo~\cite{CNS,Sakemi}. These proposed experiments motivate us to carry out this study further 
to improve the accuracies in the $R$ and $S$ calculations, so that they can be combined with the experimental results, when available, to put 
stringent bounds on $d_e$ and $C_N^{S-PS}$.
 
In this work, we focus on the properties of $^{210}$Fr on which the EDM experiment at the University of Tokyo has been planned. Fr has the highest 
predicted EDM enhancement factors out of all paramagnetic systems on which EDM experiments are currently being performed. Its projected 
sensitivity ($\vert d_e \vert \sim10^{-29} e$\, cm) is about two orders of magnitude better than the limit from $^{205}$Tl~\cite{Wundt,HaradaFPUA}. 
Furthermore, many isotopes of Fr can be
prepared~\cite{Kawamura1}, and therefore it would be possible in principle to carry out separate EDM experiments. This allows for the detection of 
signatures of CP-violating sources other than the eEDM, in particular, the S-PS interaction. The linearity of the couplings is quite similar 
for all the systems. To offset this linear dependence and to determine the source of the atomic EDMs of paramagnetic systems, several isotopes can be 
used to vary the nucleon number, which will change the S-PS coupling coefficient. Another potentially important point is that the S-PS interaction is 
more sensitive to the size of the nucleus than the electron EDM interaction, due to the contact nature of the former interaction. Currently, the most
sensitive limits for the eEDM and the S-PS coupling constant come from the paramagnetic molecule, ThO, but results from measurements and 
calculations on different Fr isotopes will also be necessary to determine both the eEDM and the S-PS interaction coupling constant. The latter 
quantity must therefore be considered when performing EDM studies in a comprehensive manner to gain insights into BSM physics, which has its own 
sensitivity to different kinds of models. So far, there have been fewer studies on the S-PS interaction than on the eEDM. The present work 
aims to provide calculations and analyses of contributions of both the eEDM and the S-PS interaction to the EDM of the ground state of $^{210}$Fr.

\section{Theory}

\subsection{BSM scenarios of eEDM and S-PS e-N interaction}

\begin{figure}[t]
\centering
\includegraphics[width=4.0cm]{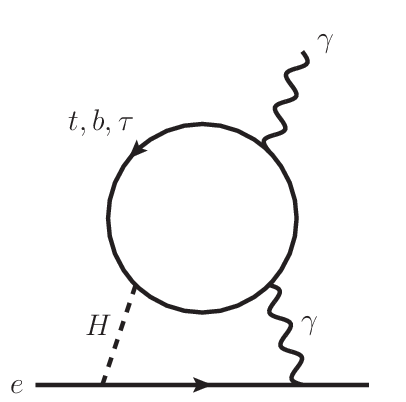}
\caption{Two-loop level contribution (so-called Barr-Zee type diagram) to the electron EDM in extended Higgs models with nonstandard Yukawa interactions.
}
\label{fig:electron_EDM_Barr-Zee}
\end{figure}

The relativistic interaction of the eEDM with the electromagnetic field $F_{\mu \nu}$ is given by 
\begin{equation}
{\cal L}_{\rm eEDM} = -d_e \frac{i}{2} \bar \psi_e \sigma^{\mu \nu} \gamma_5 \psi_e F_{\mu \nu} ,
\end{equation}
and the S-PS e-N interaction is given by
\begin{eqnarray}
{\cal L}_{\rm eN}^{\rm S-PS} 
&=&
- \sum_{N=p,n} C_N^{\rm S-PS} \frac{G_F}{\sqrt{2}} \bar \psi_N \psi_N \cdot \bar \psi_e i \gamma_5 \psi_e ,
\nonumber\\
&\equiv&
-\frac{1}{2} C_N^{\rm S-PS (0)} \frac{G_F}{\sqrt{2}} \bigl[ \bar \psi_p \psi_p +\bar \psi_n \psi_n \bigr] \bar \psi_e i \gamma_5 \psi_e
\nonumber\\
&&
-\frac{1}{2} C_N^{\rm S-PS (1)} \frac{G_F}{\sqrt{2}} \bigl[ \bar \psi_p \psi_p -\bar \psi_n \psi_n \bigr] \bar \psi_e i \gamma_5 \psi_e ,
\label{eq:SPSlagrangian}
\end{eqnarray}
where $G_F$ is the Fermi constant and $C_N^{\rm S-PS (0/1)}$ are suitably defined as the combinations of e-N S-PS coupling constants. To predict the 
existence of these two sources at the level of elementary particles, we begin with models in which the Higgs sector is augmented by nonstandard Yukawa interactions, such as the two-Higgs doublet model (2HDM). 
This is the most frequently studied 
model after the discovery of the Higgs boson at the LHC experiment \cite{Aad:2012tfa,Chatrchyan:2012xdj,Aad:2015zhl}. 
Under this extension, the EDM of the electron is generated at the one-loop level, but it is very small due to the Yukawa couplings.
The leading contribution arises instead, at the two-loop level with heavy fermion loops, often called Barr-Zee type diagrams (see Fig. 
\ref{fig:electron_EDM_Barr-Zee}) \cite{Barr:1990vd,Jung:2013hka,Cirigliano:2016njn,Brod:2018lbf,He:2020suf,Chen:2020soj,FileviezPerez:2020gfb,
Altmannshofer:2020shb}, whose leading contribution is 
\begin{eqnarray}
d_e &= & \frac{\alpha_{\rm em} e }{8 \pi^3 m_\tau} [ Y_e^P Y_\tau f (m_\tau^2 / m_H^2) + Y_e Y_\tau^P g(m_\tau^2 /m_H^2) ] \nonumber\\
&& + \frac{\alpha_{\rm em} e }{6 \pi^3 m_t} [  Y_e^P Y_t f (m_t^2 / m_H^2) + Y_e Y_t^P g(m_t^2 /m_H^2) ] \nonumber\\
&& + \frac{\alpha_{\rm em} e }{24 \pi^3 m_b} [Y_e^P Y_b f (m_b^2 / m_H^2) + Y_e Y_b^P g(m_b^2 /m_H^2) ] \nonumber\\
&\simeq & [ 2.1 \times 10^{-21} Y_e^P Y_\tau + 2.5 \times 10^{-21} Y_e Y_\tau^P \nonumber\\
&& + 4.5 \times 10^{-21} Y_e^P Y_t + 6.4 \times 10^{-21} Y_e Y_t^P \nonumber\\
&& +1.0 \times 10^{-21} Y_e^P Y_b +1.3 \times 10^{-21} Y_e Y_b^P ] \, e \, {\rm cm},
\label{eq:HDMBZ}
\end{eqnarray}
where $f(\tau) \equiv \frac{\tau}{2} \int_0^1 dx \frac{1-2x(1-x)}{x(1-x) -\tau}\ln\frac{x(1-x)}{\tau}$ and $g(\tau) \equiv \frac{\tau}{2} \int_0^1 
dx \frac{1}{x(1-x) -\tau}\ln\frac{x(1-x)}{\tau}$. Here $Y_f$ and $Y_f^P$ are the scalar and pseudoscalar couplings of the Yukawa interaction between 
the fermion $f$ and the lightest Higgs boson $H$, respectively. The Yukawa interaction Lagrangian is defined as
\begin{equation}
{\cal L}_Y = -Y_f H \bar \psi_f \psi_f -Y_f^P H \bar \psi_f i \gamma_5 \psi_f .
\end{equation}
We note that, in 2HDM, $Y_e, Y_e^P,Y_\tau, Y_\tau^P, Y_b$ and $Y_b^P$ are proportional to $\tan \beta$, the ratio between the vacuum expectation values of the two Higgs 
doublets, while $Y_t$ and $Y_t^P$ are proportional to $\cot \beta$. From Eq. (\ref{eq:HDMBZ}), we see that the expected experimental sensitivity of the
Fr atom, $\delta d_e \sim 10^{-29}e$ cm could probe the relative CP phase between the Yukawa interactions of the top quark and the electron at the 
level of $O(10^{-2})$, assuming that the mixing between the lightest and the pseudoscalar Higgs bosons is maximal. If $\tan \beta$ is large, the 
Yukawa interactions of the bottom quark or the $\tau$ lepton may be accessed. We also note that the term with $Y_b^P$ may interfere with that of 
$Y_t^P$ in Eq. (\ref{eq:HDMBZ}) if $\tan \beta $ is ${\cal O}(1)$.

\begin{figure}[b]
\centering
\includegraphics[width=4.0cm]{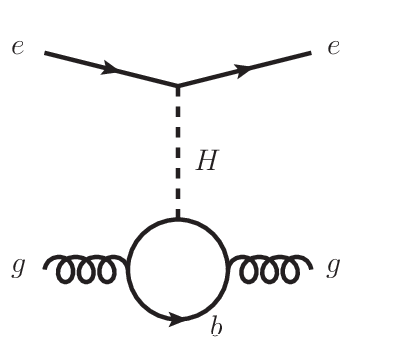}
\caption{ One-loop level contribution to the CP-odd electron-gluon interaction in extended Higgs models with nonstandard Yukawa interactions. 
The external gluons are integrated out to give 
the S-PS e-N interaction.}
\label{fig:SPS_eg}
\end{figure}

Considering the bottom quark, we note that a sizable CP-odd e-N S-PS interaction is generated through the CP-odd electron-gluon interaction 
(see Fig. \ref{fig:SPS_eg}) that gives \cite{Barr:1991yx,Barr:1992if,Cheung:2019bkw}
\begin{equation}
C^{\rm S-PS (0)}_N = \frac{\sqrt{2} \alpha_s}{12 \pi m_b G_F} \langle N |G_{\mu \nu}^a G^{\mu \nu}_a  | N \rangle \frac{Y_e^P Y_b}{m_H^2},
\end{equation}
where $-\frac{\alpha_s}{12 \pi} \langle N |G_{\mu \nu}^a G^{\mu \nu}_a  | N \rangle = 50 \pm 5$ MeV \cite{Yamanaka:2017mef,Yanase:2018qqq} and 
$m_H = 125$ GeV \cite{Aad:2015zhl}. In paramagnetic systems, this yields an ``equivalent'' contribution to the eEDM by the relation 
$\frac{S}{R} C^{\rm S-PS(0)}_N = -1 \times 10^{-21}Y_e^P Y_b\, e$ cm, which is quite comparable to the Barr-Zee type contribution (Eq. \ref{eq:HDMBZ}).
This implies that determining the sign and the precise value of $S/R$ is extremely important, since the eEDM and the equivalent one generated by 
$C^{\rm S-PS(0)}_N$ interfere destructively. To evaluate the bottom quark contribution in models with nonstandard Yukawa interactions, a precision of at least 10\% will be required for $R$ 
and $S$ as the cancellation occurs at the first significant digit itself. 
We note that this cancellation does not work with the top quark, for which the effect of the Barr-Zee type diagram exceeds that of the CP-odd e-N interaction by two orders of magnitude ($\frac{S}{R} C^{\rm S-PS(0)}_N = -3 \times 10^{-23}Y_e^P Y_t\, e$ cm).
There are also other contributions to the eEDM in this class of models \cite{Leigh:1990kf,Kao:1992jv,BowserChao:1997bb,Abe:2013qla,Egana-Ugrinovic:2018fpy}, and scenarios where the eEDM cancels were also previously 
proposed \cite{Bian:2014zka,Kanemura:2020ibp}.

\begin{figure}[t]
\centering
\includegraphics[width=6.0cm]{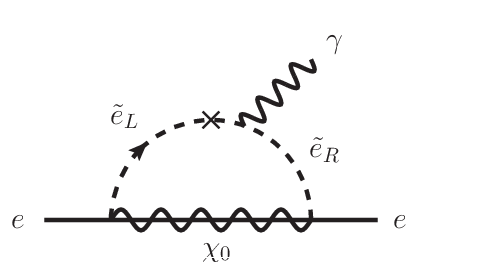}
\caption{ Example of a one-loop level diagram contributing to the eEDM in supersymmetric models. The neutralino and the selectrons are 
denoted by $\chi_0$ and $\tilde e_{L,R}$, respectively.}
\label{fig:electron_EDM_SUSY_one-loop}
\end{figure}

Supersymmetry (SUSY) in the context of particle physics provides a wide range of possibilities for BSM physics. It is a well-known fact that the CP 
violating SUSY phases are strongly constrained by the EDM \cite{Ellis:2008zy,Nakai:2016atk,Cesarotti:2018huy}. In general SUSY models predict a
large eEDM through one-loop level diagrams (see Fig. \ref{fig:electron_EDM_SUSY_one-loop}) \cite{Ellis:1982tk,Dugan:1984qf,Pospelov:2005pr}. This 
leads to
\begin{eqnarray}
d_e &\approx & \frac{e^3 }{16\pi^2 M_{\rm SUSY}^2} \Biggl[ \frac{1}{12 \cos \theta_W} \theta_{A_e}
+\Biggl( \frac{5}{\sin \theta_W} +\frac{1}{\cos \theta_W} \Biggr) \frac{\theta_\mu \tan \beta}{24} \Biggr],
\end{eqnarray}
and, assuming supersymmetric particles of mass 1 TeV, the CP phase of the slepton-Higgs trilinear coupling (the so-called $A$-term) of the first 
generation will be constrained to about $10^{-3}$ with the expected experimental data for Fr. The CP phases of the $A$-terms of the charged sleptons 
of the second and third generations, and even those of squarks may be constrained because they also contribute to the electron EDM through the two-loop
level diagrams \cite{Chang:1998uc,Yamanaka:2012ia}. The above phases are already constrained by the ThO molecule experiment \cite{ACME}, and form the 
so-called ``SUSY CP problem". There are attempts to avoid the SUSY CP problem in the most natural way, by decoupling the sfermions with very heavy mass
(split-SUSY) \cite{ArkaniHamed:2004yi}. In this scenario, several CP phases of the gaugino sector are still left, and they contribute to the eEDM at 
the two-loop level, mainly through the Barr-Zee type diagrams. The complex phases of the gaugino mass terms are already constrained at the level of 
$10^{-2}$, like the CP phase of the 2HDM. If we extend SUSYs with the R-parity violation, the R-parity violating interactions may also generate an EDM 
for Fr, but they will be constrained as well. In this case, there are many unknown couplings which may interfere with each other, and in particular, 
different couplings contribute to the eEDM \cite{Yamanaka:2012hm,Yamanaka:2012ep} and to the CP-odd e-N interaction \cite{Herczeg:1999me,
Yamanaka:2012zy}. Generally, SUSY models have many free parameters so multidimensional analyses of several experimental measurements of EDMs of 
different systems are needed to determine the unknown couplings \cite{Ibrahim:1998je,Ellis:2010xm,Yamanaka:2014nba}, and their linear independence on 
$R$ and $S$ will also be extremely important to disentangle the BSM contributions.

\begin{figure}[t]
\centering
\includegraphics[width=4.0cm]{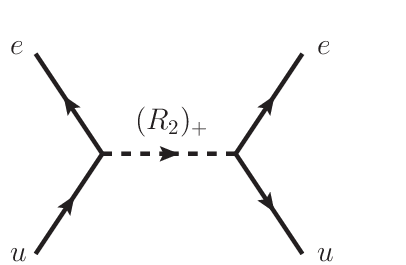}
\caption{Tree level leptoquark exchange diagram contributing to the S-PS interaction.}
\label{fig:SPS_leptoquark}
\end{figure}

Another candidate that lends itself well for EDM studies of paramagnetic systems in the context of BSM physics is the leptoquark model 
\cite{Geng:1990gr,He:1992dc,Dekens:2018bci,Herczeg:2003ag,Fuyuto:2018scm,Altmannshofer:2020ywf}. By assuming the following Yukawa interaction for the 
scalar leptoquark $(R_2)_+$ with mass $m_{\rm LQ}^2$ \cite{Herczeg:2003ag}, we get
\begin{eqnarray}
{\cal L}_{\rm LQ} = - (R_2)_+ [h'_{2L} \bar u P_L e + h'_{2R} \bar u P_R e] +{\rm h.c.},
\end{eqnarray}
the induced eEDM is given by \cite{Fuyuto:2018scm}
\begin{eqnarray}
d_e &= & -\frac{3 e m_u {\rm Im}[h'_{2L} h'^*_{2R} ] }{32\pi^2 m_{\rm LQ}^2} \Biggl[ -\frac{1}{3} + \frac{2}{3} \ln \frac{m_u^2}{m_{\rm LQ}^2}\Biggr],
\end{eqnarray}
while the CP-odd e-N coupling generated by the t-channel leptoquark exchange (see Fig. \ref{fig:SPS_leptoquark}) is given by \cite{Yanase:2018qqq}
\begin{equation}
\frac{G_F}{\sqrt{2}} C_N^{\rm S-PS(0)} = \langle N | \bar u u | N \rangle \frac{{\rm Im}[h'_{2L} h'^*_{2R} ]}{4 m_{\rm LQ}^2}.
\end{equation}
Here the contribution of the CP-odd e-N interaction is larger than that of the eEDM by two orders of magnitude. This analysis depends on the
value of the pion-nucleon sigma term $\sigma_{\pi N} \equiv \frac{m_u+ m_d}{2} \langle N |\bar uu + \bar dd | N \rangle$ whose value obtained from 
lattice QCD ($\sigma_{\pi N} \approx 30$ MeV) \cite{Yang:2015uis,Durr:2015dna,Bali:2016lvx,Yamanaka:2018uud,Alexandrou:2019brg,Borsanyi:2020bpd} and 
phenomenological extractions ($\sigma_{\pi N} \approx 60$ MeV) \cite{Alarcon:2011zs,Hoferichter:2015dsa,Yao:2016vbz,RuizdeElvira:2017stg,
Friedman:2019zhc,Huang:2019crt,Ma:2020kjz} are currently not in agreement. We also note that the leptoquark model contributes to the 
tensor-pseudotensor type CP-odd e-N interaction, so constraining the CP phases of leptoquark models also requires the precise calculation of the 
corresponding atomic level coefficient.

\begin{figure}[t]
\centering
\includegraphics[width=5.0cm]{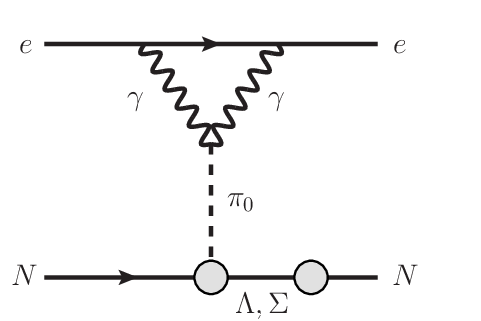}
\caption{Leading two-photon exchange contribution to the S-PS interaction in the SM. The grey blobs are strangeness changing interactions generated by the CKM matrix.}
\label{fig:SPS_SM}
\end{figure}

We  briefly touch upon the SM contribution to the paramagnetic systems induced by the CP phase of the CKM matrix. As mentioned in the Introduction, the electron 
EDM is small ($d_e \sim 10^{-39} e$ cm) \cite{Yamaguchi:2020eub,Yamaguchi:2020dsy}. The leading contribution to the paramagnetic systems is actually 
from the S-PS interaction (see Fig. \ref{fig:SPS_SM}), with its coupling estimated as \cite{He:1992dc,Pospelov:2013sca,Yamanaka:2015ncb,Yamanaka:2017mef}
\begin{equation}
C_N^{\rm S-PS(0,1)} \sim {\cal O}(10^{-17}).
\end{equation}
This generates an atomic EDM, which is larger than that induced by the eEDM by one or two orders of 
magnitude at the atomic level, but precise values of both quantities will be required to determine the SM contribution due to the large 
theoretical uncertainty. Needless to say, this is much smaller than the TeV scale BSM effects presented in this section with the natural sizes of CP phases and 
coupling constants.

\subsection{The $R$ and $S$ factors}

 The Hamiltonians describing the eEDM and S-PS e-N interactions in atomic systems are given by 
\begin{eqnarray}
 H^{eEDM} = -d_e \beta \vec \Sigma \cdot {\cal \vec E}^{int} ,
\end{eqnarray}
and 
\begin{eqnarray}
 H^{S-PS} = i \frac{G_F}{\sqrt{2}} C_N^{S-PS(0)} A \beta \gamma_5 \rho(r) ,
\end{eqnarray}
respectively, where ${\cal \vec E}^{int}$ is the total internal electric field experienced by the electrons in an atom, $A$ is the atomic mass number, 
$\rho(r)$ is the nuclear density and $\vec \Sigma = \begin{pmatrix} \vec \sigma & 0 \\ 0 & \vec \sigma \end{pmatrix}$ is the four-component 
spinor with Pauli matrix $\vec \sigma$ as its elements. We have assumed a Fermi charge distribution for the nuclear density.

An atomic state ($\vert \Psi_v^{(0)} \rangle$), in the absence of any external interaction (denoted with superscript $0$), is an eigenstate
of the atomic Hamiltonian ($\hat{H}_0$), and can be written as
\begin{eqnarray}
 \hat{H}_0 \vert \Psi_v^{(0)} \rangle = E_v^{(0)} \vert \Psi_v^{(0)} \rangle,
\end{eqnarray}
where $E_v^{(0)}$ is the corresponding energy. In our calculations, we consider first the Dirac-Coulomb (DC) interactions in $\hat{H}_0$
then add Breit and QED interactions systematically to assess the importance of the higher-order relativistic effects. 
If $H^{eEDM}$ or $H^{S-PS}$ is considered as the first-order perturbation, the new atomic state wave function ($\vert \Psi_v \rangle$)
can be written as
\begin{equation}
\vert \Psi_v \rangle \approx \vert \Psi_v^{(0)} \rangle + \lambda \vert \Psi_v^{(1)} \rangle, \label{eq:perturbedWF}
\end{equation}
where $\vert \Psi_v^{(1)} \rangle$ is the first-order perturbed wave function with the perturbation parameter $\lambda \equiv d_e$ (or
$\lambda \equiv C_N^{S-PS(0)}$) when the eEDM (or S-PS) interaction Hamiltonian is considered after defining the interaction Hamiltonian $H'=H^{eEDM}/d_e$ (or 
$H'=H^{S-PS}/C_N^{S-PS(0)}$). In our approach, we obtain $\vert \Psi_v^{(1)} \rangle$ by solving 
\begin{eqnarray}
 (\hat{H}_0 -  E_v^{(0)}) \vert \Psi_v^{(1)} \rangle = (E_v^{(1)}-H') \vert \Psi_v^{(0)} \rangle,
\end{eqnarray}
where $E_v^{(1)}$ is the first-order perturbed energy and is equal to zero here owing to the odd-parity nature of $H'$. An important feature 
of solving for $\vert \Psi_v^{(1)} \rangle$ in this work is that it circumvents the problem of excluding high-lying intermediate states in 
the sum-over-states approach, thereby implicitly including all the intermediate states in the calculations.

The atomic EDM is given by 
\begin{eqnarray}
D_a &=& 2 \lambda \frac{\langle \Psi_v^{(0)} \vert D_z \vert \Psi_v^{(1)} \rangle}{\langle \Psi_v^{(0)} \vert \Psi_v^{(0)} \rangle} ,
\label{eq:aEDMexpr}
\end{eqnarray}
with the electric dipole (E1) operator ${\vec D}$ so that $R$ and $S$ are defined as
\begin{eqnarray}
R = D_a / \lambda \ \ \ \text{and} \ \ \ S = D_a  / \lambda ,
\end{eqnarray}
for $\lambda=d_e$ and $\lambda=C_N^{S-PS(0)}$, respectively. It is worth noting that ${\vec D} \equiv e {\vec r} + d_e \beta {\vec \Sigma}$ and ${\vec D} \equiv e {\vec r}$ in 
the case of eEDM and S-PS interactions, respectively. 

\subsection{RCC theory of $R$ and $S$}

The relativistic coupled-coupled cluster (RCC) theory has been employed to determine $R$ and $S$ \cite{Fr1,Nataraj,Shee}, which is arguably the current gold 
standard for relativistic many-body theory to determine atomic wave functions of the closed-shell and one-valence atoms and molecules. In the RCC theory,
the unperturbed and perturbed wave functions of one-valence system like Fr atom are expressed as (e.g. see \cite{Nataraj,Fr1})
\begin{eqnarray}
 \vert \Psi_v^{(0)} \rangle = e^{T^{(0)}} \left \{ 1 + S_v^{(0)} \right \} \vert \Phi_v \rangle ,
\end{eqnarray}
and 
\begin{eqnarray}
 \vert \Psi_v^{(1)} \rangle = e^{T^{(0)}} \left \{ T^{(1)} (1+ S_v^{(0)}) + S_v^{(1)} \right \} \vert \Phi_v \rangle ,
\end{eqnarray}
respectively. In the above expressions, $T$ and $S_v$ operators are classified as core and core with valence correlation excitation operators with superscripts
0 and 1 representing the unperturbed and perturbed interactions, respectively, and $\vert \Phi_v \rangle$ is the reference state function, which is the 
Dirac-Hartree-Fock (DHF) wave function in the present work. Detailed procedures to evaluate amplitudes of $T^{(0/1)}$ and $S_v^{(0/1)}$ operators can be found 
elsewhere (e.g. see \cite{Nataraj,Fr1}). A remarkable virtue of the RCC theory is that it enables the evaluation of the many-body effects to all-orders.

Substituting the above expressions in Eq. (\ref{eq:aEDMexpr}), we have
\begin{eqnarray}
 \frac{D_a}{\lambda}  = 2 \frac{\langle \Phi_v \vert \left \{ 1+ S_v^{(0)\dagger} \right \} \bar{D} \left \{ T^{(1)}( 1+ S_v^{(0)}) + S_v^{(1)}\right \} \vert \Phi_v \rangle}
 {\langle \Phi_v \vert \left \{ 1+ S_v^{(0)\dagger} \right \} \bar{N} \left \{ 1+ S_v^{(0)}  \right \} \vert \Phi_v \rangle}, \nonumber 
\end{eqnarray}
where $\bar{D}=e^{T^{(0)\dagger}} D e^{T^{(0)}}$ and $\bar{N}=e^{T^{(0)\dagger}} e^{T^{(0)}}$. Clearly, the above expression corresponds to $R$ and $S$ for 
$\lambda=d_e$ and $\lambda=C_N^{S-PS(0)}$, respectively. We evaluate the numerator part first then extract out the contribution due to the normalization of wave function 
(given as `Norm') separately. The series of terms representing $\bar{D}$ and $\bar{N}$ are both non-terminating. However, we had ignored terms beyond quadratic 
power in the RCC operators in our previous works \cite{Nataraj,Fr1}. Here, we have evaluated them self-consistently including contributions from higher order non-linear terms
with tolerance values of $10^{-8}$ for $R$ and $S$. In the present work, we have considered singles and doubles excitations (RCCSD method) in our RCC theory. A 
subset of triple excitation terms were also included perturbatively in the present work.
\begin{table}[t!]
\caption{DHF and RCC results for the EDM enhancement factors $R$, and the S-PS coupling constants $S$ of the atomic EDM of $^{210}$Fr  
using the DC Hamiltonians, calculated using the RCCSD method. $S$ values are given in units of $10^{-18}e$ cm. These values are compared 
against previous calculations by Ref.~\cite{Fr1}.}
\begin{center}
 \begin{tabular}{lccc} 
 \hline
 \hline
 & DHF & RCC & Other calculations \\
 \hline
 $R$ & 727.24 & 812.19 & 721.21~\cite{Fr1} \\
 &   & & 894.93~\cite{Fr1}  \\
 $S$ & 9.5309 & 10.621 &  $10.862  \pm 0.3$ \cite{skrip}  \\
 \hline
 \hline
\end{tabular}

\label{tb:DC-210-212}
\end{center}
\end{table}

\section{Results and Discussion}

Our work aims to surpass past results~\cite{Fr1, Fr2, Fr3, skrip}, by using an improved RCC method that addresses some of the shortcomings of the 
previous RCC calculation by Mukherjee et al.~\cite{Fr1}. The advances that have been made include an improved basis set (see Ref. \cite{arXiv}) and the 
inclusion of terms that were omitted in previous calculations~\cite{Fr1,Fr2} due to computational cost limitations. In view of the recent progress in 
ongoing Fr EDM experiments~\cite{HaradaFPUA}, evaluating an improved theoretical result is of importance in yielding limits for the eEDM and the
S-PS interaction coupling constants. High-performance computing is utilized in these calculations to include as many 
terms as possible for improved accuracy, as well as inclusion of correction terms due to physical effects, such as the Breit interaction and QED 
effects, which were not considered in previous Fr EDM electron enhancement factor calculations. Contributions from the valence triple excitation
terms are also included in the present work after evaluating them perturbatively. Table~\ref{tb:DC-210-212} shows DHF and RCC results for $R$ and $S$ 
of $^{210}$Fr, evaluated using the DC Hamiltonian. After adding the contributions of the Breit interaction, the QED effects and perturbative 
triple excitations (see Ref. \cite{arXiv} for details), the following results are obtained along with the uncertainties
\begin{eqnarray}
R &=& 799 \pm 24 \\
\text{and} \ \ \ S &=& (10.5 \pm 0.3) \times 10^{-18} \ \ e~\text{cm} .
\end{eqnarray}
The Breit and QED corrections follow almost similar trends in the determination of both $R$ and $S$. It would be in order at this juncture to comment on 
our approach to the error estimation for $R$ and $S$. Both of the quantities are proportional to the atomic EDM given by Eq. (\ref{eq:aEDMexpr}), 
which shows that they depend on the unperturbed and the first-order perturbed wave functions. The later quantity, i.e. the first-order perturbed 
wave function, can be written as
\begin{eqnarray}
 \vert \Psi_v^{(1)} \rangle = \sum_{I\ne v} \vert \Psi_I^{(0)} \rangle \frac{\langle \Psi_I^{(0)}\vert  H_{CPV} \vert \Psi_v^{(0)}\rangle}{E_v^{(0)}-E_I^{(0)}},
\end{eqnarray}
where $H_{CPV}$ could be either $H^{eEDM}$ or $H^{S-PS}$. Thus, Eq. (\ref{eq:aEDMexpr}) can be rewritten as
\begin{eqnarray}
 D_a = \sum_{I\ne v} \frac{\langle \Psi_v^{(0)}\vert D \vert \Psi_I^{(0)} \rangle \langle \Psi_I^{(0)}\vert  H_{CPV} \vert \Psi_v^{(0)}\rangle}{E_v^{(0)}-E_I^{(0)}}.
\label{eqnsum}
 \end{eqnarray}
For Fr, the unperturbed state $\vert \Psi_I^{(0)} \rangle$ is the ground state, i.e. 7S state. By far, the dominant contribution to Eq. (\ref{eqnsum})
comes from the $7P_{1/2}$ state. Therefore, a reasonable estimate of error in $D_a$, and hence $R$ and $S$, can be found by determining and adding 
in quadrature the uncertainties of the E1 matrix element $\langle 7P_{1/2} \vert D \vert 7S \rangle$, the matrix element of the EDM 
interaction Hamiltonian $\langle 7P_{1/2 }\vert  H_{CPV} \vert 7S\rangle$ and the excitation energy $E_{7S}^{(0)} - E_{7P_{1/2}}^{(0)}$. Our calculated
values and their comparison with the experimental values of the above E1 matrix element and excitation energy can be found in Ref. \cite{arXiv}. the 
error in the EDM matrix element is obtained by taking the deviation of our calculated value of the geometric mean of the magnetic dipole hyperfine 
constants ($A_{hf}$) of the $7S$ and $7P_{1/2}$ states from its measured values. The underlying reason for this is that the evaluation of the EDM 
interaction matrix elements that we have considered in the present work and the $A_{hf}$ values are sensitive to the nuclear region. 

Using the approach outlined above and from the deviation of our calculated values from their experimental values as shown in Ref. \cite{arXiv}, the 
uncertainties obtained in $R$ and $S$ for $^{210}$Fr from the $7P_{1/2}$ state are 2.5\% for both the quantities. From the trends observed in the contributions to $R$ and 
$S$ from other high-lying $P_{1/2}$ states as well as the accuracies of the E1 matrix elements involving these states and the ground $7S$ state, 
and the excitation energies of these $P_{1/2}$ states, we infer that the overall uncertainties in $R$ and $S$ for $^{210}$Fr are about 3\%. The error
analyses reported in the previous calculations of $R$ and $S$ were not based on the uncertainties in the atomic properties that are mentioned above.

As mentioned earlier, there are three major differences between our present and previous result from Ref. \cite{Fr1}. In this work (i) we have increased the
size of our single particle Gaussian basis and improved its quality compared to our previous work. (ii) We have included RCC amplitudes corresponding 
to all multipoles of the Coulomb interaction and not a selected set of amplitudes that was used in our previous work. (iii) The Breit interaction, 
approximate QED and perturbative triple excitations have been included unlike in our previous calculation. In contrast to our full {\it ab initio}
calculation, Byrnes et al.~\cite{Fr2} have carried out a part of {\it ab initio} and a part semi-empirical calculations. Furthermore, the latter 
evaluates $R$ by summing over a set of intermediate $P_{1/2}$ states. On the other hand, our calculation implicitly accounts for all the intermediate 
states in the configuration space that has been considered in our work. They also estimate the errors in excitation energies, E1 transition amplitudes,
and the EDM matrix elements, and reported a total error of 5\%. Our calculations provide a comprehensive analysis of the errors in $R$ and $S$. The 
result for $R$ reported by Sandars \cite{Fr3} has been obtained by a one-body potential, and it is therefore very approximate. Neither Byrnes et al.
nor Sandars have considered higher-order relativistic effects like the Breit interaction and QED effects. The only calculation of $S$ available for 
Fr in the literature is by Skripnikov et al. \cite{skrip}. This work is based on a variant of RCC theory. Its reported value $S  = (10.862  \pm 0.3)
\times 10^{-18} \ e$ cm  differs from our calculation of $S$ by 3.4\% and its quoted error is about 3\%. It is not clear to us how they have estimated
the error since they have not calculated the relevant E1 matrix elements and $A_{hf}$ values. They have evaluated $S$ as a second derivative of the 
ground state energy with respect to the external field and $C_N^{S-PS(0)}$, rather than an expectation value as we have. 
The other possible reason for the discrepancy between the two calculations is the treatment of the higher-order relativistic effects. Skripnikov et al. have considered the Gaunt term of the Breit interaction, which in most cases gives the leading contribution. But they have not considered the retardation part of this interaction. We have accounted for both the terms in the Breit interaction.

\begin{figure}[t]
\centering
\includegraphics[width=8.6cm]{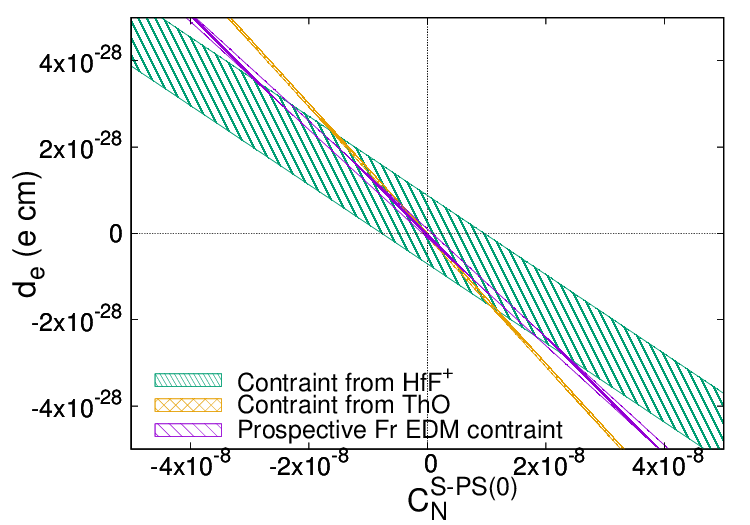}
\caption{Experimental constraints of $d_e$ versus $C_N^{\rm S-PS(0)}$ with the prospective sensitivity of the EDM of Fr atom. The existing data given 
by the experiments using ThO \cite{ACME} and HfF$^+$ \cite{Cairncross:2017fip} molecules are also shown.}
\label{fig:CSPSvsEDM}
\end{figure}

\section{Analysis of the elementary level CP violation}

We analyze now the implications of our results bearing in mind the projected sensitivity of the Fr EDM experiment. The expected experimental 
accuracy is often expressed in terms of the electron EDM, as $\delta d_e \sim 10^{-29}e$ cm. As we have emphasized earlier, measurements on paramagnetic 
systems are also sensitive to the S-PS interaction, so only its linear combination with the electron EDM can be constrained. The actual inequality is then
\begin{equation}
\left| \, d_e + \frac{S}{R} C_N^{S-PS(0)}  \right| < 10^{-29}e \, {\rm cm} .
\end{equation}
The measurement of the EDM of $^{210}$Fr with an accuracy of $\delta d_e \sim 10^{-29}e$ cm, combined with that of the current experimental results 
from $^{232}$ThO \cite{ACME} and HfF$^+$ \cite{Cairncross:2017fip} may determine the splitting between $d_e$ and $C_N^{S-PS(0)}$ by roughly 10\% 
(see Fig. \ref{fig:CSPSvsEDM}). The sensitivity to this splitting may also be enhanced by the deformation of the nucleus, as in the case
of $^{224}$Fr, and this requires a new calculation since the nuclear density profile changes.

Measurement on several isotopes may ultimately provide the sensitivity to the isospin splitting of $C_N^{S-PS}$, i.e. disentangling $C_p^{S-PS}$ and 
$C_n^{S-PS}$. However, we note that the isospin splitting is suppressed at the quark level by the close values of the quark scalar densities of the
nucleons. At the quark level, $C_N^{S-PS}$ is proportional to the nucleon scalar density $\langle N| \bar qq |N\rangle \sim 10$ ($q=u,d$, 
$N=p,n$) \cite{Yamanaka:2017mef}, extracted from lattice QCD calculations \cite{Yang:2015uis,Durr:2015dna,Bali:2016lvx,Yamanaka:2018uud,
Alexandrou:2019brg,Borsanyi:2020bpd} and phenomenology \cite{Alarcon:2011zs,Hoferichter:2015dsa,Yao:2016vbz,RuizdeElvira:2017stg,Friedman:2019zhc,Huang:2019crt},
while the isospin splitting $|\langle N| \bar uu - \bar dd |N\rangle | \sim 1$ \cite{Bali:2014nma,Yang:2015zja,Dragos:2016rtx,Yoon:2016jzj,
Yamanaka:2018uud,Egerer:2018xgu,Hasan:2019noy,Harris:2019bih,Alexandrou:2019brg}, which is in view of the theoretical uncertainty, not observable.
This affirms that the quark level isospin splitting cannot be discerned by the measurement of the EDM of several isotopes of Fr.

\section{Conclusion}

Improved RCC calculations of the EDM of $^{210}$Fr arising due to the eEDM and the S-PS interaction were performed, and the results obtained 
are 
\begin{equation}
D_a= 799 \, d_e + 10.5 \times 10^{-18} \, C_N^{S-PS(0)} e\, {\rm cm} ,
\end{equation}
with an estimated error of about 3\% for both cases. The former value is about 11\% smaller than the result from a previous calculation of this 
quantity using an approximate RCCSD method~\cite{Fr1}. This difference can be attributed to the fact that the various approximations and shortcomings
in the previous calculation were addressed in this work, such as the improvement of both the size and quality of the basis functions used, and the 
inclusion of cluster amplitudes for all multipoles and computation of the nonlinear RCC terms using a self-consistent approach. We emphasize that we
have outlined the method of error evaluation more comprehensively than what is given in previous calculations on the same atom. A detailed analysis of 
the many-body effects contributing to $R$ and $S$ for $^{210}$ Fr has been made. This has shed light on the many-body physics involved in this complex phenomenon involving electromagnetic and CP 
violating interactions. Our work also includes corrections due to the Breit interaction and QED effects, as well as contributions due to perturbative 
triple excitations, which previous Fr EDM calculations have not included. The reasons for the relatively small discrepancy between our result for $S$
and a previous calculation of the same quantity have been discussed.

Using the value of $R$ that we calculated for $^{210}$Fr and quantities related to a laser cooled experiment on this isotope, we obtained as mentioned
before, a sensitivity, $\delta d_e \sim 10^{-29}e$ cm, which will permit us to probe many CP violating phases associated with the well-known TeV scale
BSM physics at the same level as in ThO. It is of particular interest to note that our precise results will enable us to analyze the bottom quark 
contribution in models where the Higgs sector is extended by nonstandard Yukawa interactions in which the contributions of the electron EDM and the S-PS electron-nucleon interaction will strongly cancel. The precision of
3\% that we could achieve is clearing the minimal required one of 10\%. We also point that our results of the Fr atom, combining with calculations for
other paramagnetic systems, will allow us to disentangle the CP violation generated by different elementary level sources, which is the case in BSM
candidates involving many CP phases such as SUSY. Further accurate evaluations of the coefficient of the tensor-pseudotensor type CP-odd e-N 
interaction and the pion-nucleon sigma term would also be required to disentangle the CP violating effects in the models where the contributions from
$d_e$ and $C_N^{S-PS(0)}$ interfere destructively. The dependence of the sensitivity of the EDM of the Fr atom on the nucleon number may also vary significantly,
and therefore a judicious choice of isotopes would be necessary. These are important subjects for future investigations.

The difference between the results of our and those of previous calculations suggests that it is necessary to continue to perform these 
calculations to higher levels of accuracy, for a reliable appraisal of the upper limit of the magnitude of the electron EDM in conjunction with future
experimental results. For a comprehensive study of BSM physics, it is necessary to conduct EDM searches on multiple candidates. In this respect, the 
continuation of efforts for EDM measurement in atomic systems is still important. The merit of Fr in particular as a candidate for an EDM 
search is its large size of $R$ and $S$, and the high accuracy to which it can be obtained compared to molecules of interest for EDM experiments. 
In addition, it is an ideal system in which to also probe the S-PS  coupling constants because of the wealth 
of Fr isotopes that can be produced. The theoretical work that we have reported will complement experimental efforts that are underway to observe EDM 
in Fr, in addition to providing insights into the interplay of relativistic many body effects and CP violating interactions in atoms.

\acknowledgments

The calculations in this work were performed on the TSUBAME 3.0 supercomputer at the Tokyo Institute of Technology, through the TSUBAME Grand
Challenge Program. The authors would like to thank Professor Y. Sakemi and Dr. K. Harada for providing useful information on experimental aspects of
Fr EDM and Dr. V. A. Dzuba for providing useful information related to his group's calculation of the electron EDM enhancement factor for Fr.

\end{document}